\documentclass[aps,prd,onecolumn,groupedaddress,showpacs,nofootinbib,amssymb
]{revtex4}
\usepackage[dvips]{graphicx}
\usepackage{amssymb}
\usepackage{amsmath}
\usepackage{epstopdf}
\usepackage{graphicx,,color}
\usepackage{amsfonts}
\usepackage{bm}
\usepackage{cancel}
\usepackage{comment}

\allowdisplaybreaks[4]

\begin{document}

\tolerance=5000



\title{On wormhole formation in $f(\textit{R}, \textit{T})$ gravity: \\varying Chaplygin gas and barotropic fluid}

\author{
Emilio Elizalde$^{1,2,3}$\thanks{E-mail: elizalde@ieec.uab.es}, 
Martiros Khurshudyan$^{2,3,4,5}$\thanks{Email: khurshudyan@yandex.ru, khurshudyan@tusur.ru, khurshudyan@ustc.edu.cn}}

\affiliation{
$^1$ Consejo Superior de Investigaciones Cient\'{\i}ficas, ICE/CSIC-IEEC,
Campus UAB, Carrer de Can Magrans s/n, 08193 Bellaterra (Barcelona) Spain \\
$^{2}$ International Laboratory for Theoretical Cosmology, Tomsk State University of Control Systems 
and Radioelectronics (TUSUR), 634050 Tomsk, Russia \\
$^{3}$ Research Division, Tomsk State Pedagogical University, 634061 Tomsk, Russia \\
$^{4}$ CAS Key Laboratory for Research in Galaxies and Cosmology, Department of Astronomy, University of Science and Technology of China, Hefei 230026, China \\
$^{5}$ School of Astronomy and Space Science, University of Science and Technology of China, Hefei 230026, China \\
}

\begin{abstract}
In this paper, the formation of specific static wormhole models is discussed, by assuming an $f(\textit{R}, \textit{T}) = R + 2 \lambda T$ extended theory of gravity,  $T =- \rho + P_{r} + 2P_{l}$ being the trace of the energy momentum tensor. In the first part, wormhole solutions are constructed imposing that the radial pressure admits an equation of state corresponding to a varying Chaplygin gas. Two forms for the varying Chaplygin gas are considered, namely $P_{r} = - B b(r)^{u}/\rho^{\alpha}$ and $P_{r} = -B R(r)^{m}/ \rho^{\alpha }$, respectively. In the second part, the wormhole models are constructed assuming that the radial pressure can be described by a varying barotropic fluid. In particular, $P_{r} = \omega b(r)^{v} \rho$ and $P_{r} =\hat{ \omega} r^{k} R(r)^{\eta}  \rho$ are considered, respectively, leading to two additional, traversable wormhole models. In all cases, $b(r)$ is the shape function, and $R(r)$ the Ricci scalar obtained from the wormhole metric for a redshift function equal to $1$. With the help of specific examples, it is demonstrated that the shape functions of the exact wormhole models previously constructed do obey the necessary metric conditions. The same energy conditions help reveal the physical properties of these models. A general feature is the violation of the NEC~($\rho + P_{i} \geq 0$) in terms of the radial pressure $P_{r}$ at the throat of the wormhole. For some of the models, one can satisfy the NEC at the throat while a violation of the DEC~($\rho - P_{i} \geq 0$) occurs. To summarize,  exact wormhole models can be constructed with a possible violation of the NEC and DEC at the throat of the wormhole, while being $\rho \geq 0 $. Thus, the interesting feature appears that one has a violation of the WEC~($\rho \geq 0$ and $\rho + P_{i} \geq 0$) not related to the energy density behavior (the index i, being r resp. l, indicates radial resp. lateral pressure). 

\end{abstract}

\pacs {}

\maketitle

\section{Introduction}\label{sec:INT}

Construction of exact wormhole models was always a challenging task in General Relativity~(GR), and it remains so in modified theories of the same. The study of wormhole solutions is a most active field in the area. In the recent literature, one finds various interesting studies, such as~\cite{Jawad:2016}~-~\cite{Zubair:2017}~(to mention  a few, only); in particular, for the case of modified theories of gravity, which have proved to be efficient in order to solve some of the problems of modern cosmology~(see, e.g., \cite{Nojiri:2017ncd}~-~\cite{Oikonomou:2016jjh}). It is well known that, according to the observational data, an important part of the energy content of the Universe is under the form of a dark energy, which accelerates its expansion; this bein true, provided GR is taken to describe the background dynamics~(see, for instance, \cite{Planck:2015} and \cite{Delubac:2014dt}, and references therein). In general, any specific energy source which can generate negative pressue should be counted as dark energy. There are various models of dark energy, addressing the accelerated expansion problem, well constrained by observational data~\cite{Xia:2016vnp}~-~\cite{Brevik:2017msy}~(and references therein). On the other hand, if for some of these dark energy models, some issue, for instance, the cosmological coincidence problem, cannot be solved, then one can still turn to consider non-gravitational interaction, to finally obtain a consistent picture. Discussions on this issue can be found in~\cite{Xia:2016vnp}~-~\cite{Brevik:2017msy}, and  references therein. Moreover, specific modifications of GR can be used to solve both the dark energy and the dark matter problems, and some others. 

As a consequence of the above, it is very important to analyze all possibilities, in particular, the ones involving cosmological objects, like black holes, gravastars, strange stars, and other, including of course wormholes, and also in the case of modified theories of gravity. In this context, Refs.~\cite{Astashenok:2013vza}~-~\cite{Maria:2018}, and the citations therein, are quite useful to get a general understanding on some of these subjects and on the possibilities of future developments concerning the above mentioned topics, specifically, for the case of $f(R)$ gravity. 

More about whormoles. Being those hypothetical objects able to connect asymptotical regions of a single Universe, their study in modified gravity theories can indeed provide valuable additional information on the above mentioned issues; in particular, if it could be proven that such a wormhole can actually work as a physical tunnel connecting two distinct Universes. We actually believe that they may play a crucial role for a better understanding of the quantum nature of the early Universe. However, to this end wormholes should be first detected, and their matter content should be also well constrained and understood, in order to sustain the hope that we can extract useful information from them. We should consider that even the data allowing to constrain and understand the nature and content of other objects, which may sound much more familiar, as black holes or even dark matter, is scarce and this makes the situation very complicated, not to speak about wormholes! What are the hints on the nature and content of a wormhole? Even in cosmology, when today we are in possession of relatively good observational data from different missions, we still have enormous problem to understand what is the real nature of dark matter, and many different possible candidates are being discussed, some of them quite afar from each other. We have to face an even more difficult situation, in the case considered of a wormhole. 

Different proposals on the Universe energy structure have been put forward by different theories of modified gravity. A particular example is $f(\textit{R}, \textit{T})$ gravity, with $T = \rho + P_{r} + 2P_{l}$ the trace of the energy momentum tensor. In general, one could expect that the material corrections should come either from the existence of imperfect fluids, or from quantum effects, such as particle production. The total action of such theories is generally given by~\cite{Harko:2011kv}
\begin{equation}\label{eq:Action}
S = \frac{1}{16 \pi} \int{ d^{4}x\sqrt{-g} f(\textit{R}, \textit{T}) } + \int{d^{4}x\sqrt{-g} L_{m}},
\end{equation}
where $f(\textit{R}, \textit{T})$ is an arbitrary function of the Ricci scalar $R$ and of the trace of the energy-momentum tensor $T$, while $g$ is the metric determinant, and $L_{m}$ is the matter Lagrangian density, which is related
to the energy-momentum tensor in the following way
\begin{equation}
T_{ij} = -\frac{2}{\sqrt{-g}} \left[  \frac{\partial (\sqrt{-g} L_{m}) }{\partial g^{ij}} - \frac{\partial}{\partial x^{k}} 
\frac{\partial(\sqrt{-g}L_{m})}{\partial(\partial g^{ij}/\partial x^{k})} \right].
\end{equation}
If we assume that $L_{m}$ depends on the metric components only, then
\begin{equation}
T_{ij} = g_{ij}L_{m} - 2 \frac{\partial L_{m}}{\partial g^{ij}}.
\end{equation}
Moreover, varying the action, Eq.~(\ref{eq:Action}), with respect to the metric $g_{ij}$ yields the field equations
$$f_{R}(\textit{R}, \textit{T}) \left( R_{ij} - \frac{1}{3} R g_{ij} \right) + \frac{1}{6} f(\textit{R}, \textit{T})  g_{ij}  = 8\pi G \left( T_{ij} - \frac{1}{3} T g_{ij} \right) -f_{T}(\textit{R}, \textit{T})  \left( T_{ij} - \frac{1}{3} T g_{ij} \right)$$
\begin{equation}
-f_{T}(\textit{R}, \textit{T})  \left( \theta_{ij} - \frac{1}{3} \theta g_{ij} \right) + \nabla_{i}\nabla_{j} f_{R}(\textit{R}, \textit{T}) ,
\end{equation}
with $f_{R}(\textit{R}, \textit{T}) = \frac{\partial f(\textit{R}, \textit{T}) }{\partial R}$, $f_{T}(\textit{R}, \textit{T}) = \frac{\partial f(\textit{R}, \textit{T}) }{\partial T}$ and
\begin{equation}
\theta_{ij} = g^{ij} \frac{\partial T_{ij}}{\partial g^{ij}}.
\end{equation}

Going now back to the discussion of the dark energy and dark matter components of our Universe, we find in the literature quite different possibilities for the same. One of the first, for dark energy, is under the form of a scalar field, quite successfully applied to the accelerated expanding Universe problem. There is also a way to present dark energy as a fluid. In this case, the barotropic fluid model, with a negative equation of state parameter, is maybe the simplest one, among others. There are also other attempts to present dark energy as a fluid including various parametrizations of the energy density and pressure.  On the other hand, the consideration of a fluid~(the energy source) with an equation of state which unifies dark energy and dark matter is also a very compelling approach and could significantly simplify the whole picture and the related analysis. One of such examples is the Chaplygin gas, which has a non-linear equation of state. There is a good number of papers addressing important issues of modern cosmology by using this equation of state. On the other hand, various modifications of the original form of the Chaplygin gas have been considered, aiming in particular to a better understanding of the physics behind the most recent astronomical observations. The references at the end of this paper can be used as a guide to  the possibilities mentioned above.

Previous studies on wormhole models in modified theories of gravity, assuming different matter content, reveal the viability of several different hypotheses concerning their matter content. However, since wormholes have never been observed, we cannot be certain about such hypotheses, neither in GR nor in modified theories of gravity, as has been mentioned at the beginning, already. 

Being aware of this situation, in this paper we address the wormhole formation problem in $f(\textit{R}, \textit{T}) = R +2  \lambda T$  gravity assuming two families of equations of state to describe the matter content of the wormhole. More specifically, in the first part of the paper we will consider wormhole models assuming that the radial pressure has the following form
\begin{equation}\label{eq:VCGGEN}
P_{r} = - \frac{B(r)}{\rho^{\alpha}},
\end{equation}   
where $A$ and $\alpha$ are constant, while $B(r)$ is a function of $r$. The last assumption associates the matter content of the wormhole to a varying Chaplygin gas with the radial pressure being given by Eq.~(\ref{eq:VCGGEN}). We also study wormhole formation for the case of $f(\textit{R}, \textit{T}) = R +2  \lambda T$, with $P_{r} = - B / \rho ^{\alpha}$, and find that, here, the traversable wormhole solution should be described by a function of constant shape, $b(r)$. In particular, we will find that a traversable wormholes with $b(r) = r_{0}$ is formed in such scenario. Another interesting aspect observed during the study is that $\rho \approx 0$, for these cases, what implies that $P_{r}$ and $P_{l}$ are infinite. However, when we consider $P_{r} = - B(r) / \rho ^{\alpha}$ with $B(r)$ to be either a function of the Ricci scalar, or a function of the shape function $b(r)$, we will have a wormhole solutions with finite $P_{r}$ and $P_{l}$. In other words, the consideration of a varying Chaplygin gas proposed here will allow to overcome the problems that have been reported for constant $B$. This means that there could be a deep reason for this substantial change of behavior,  and that we need to consider more complicated forms of $f(\textit{R}, \textit{T})$ in other to better understand the situation. This could indicate also, that the reasons allowing to consider matter corrections simply prevent the wormhole formation in a $B = const$ Chaplygin gas. Of course, this possibility may introduce some constrains and the need for a deeper study of the modified gravity theory of the kind $f(\textit{R}, \textit{T}) = R +2  \lambda T$, what will be discussed in a forthcoming paper, together with other relevant questions.  However, we should remember that when we assume the metric of the universe to be given by the flat Robertson-Walker metric, then the $f(\textit{R}, \textit{T}) = R +2  \lambda T$ model is equivalent to a cosmological model with an effective cosmological constant proportional to $H^{2}$. Moreover, in this particular model the gravitational coupling becomes an effective and time dependent coupling. This is of course a very important result for cosmology, which motivates us to study the impact of the considered form of $f(\textit{R},\textit{T})$ gravity on wormhole solutions and energy conditions~(see more details in~\cite{Harko:2011kv}).

In the second part of this work, we construct two more models,  assuming that the radial pressure can be described by a varying barotropic fluid. In particular, we consider the cases $P_{r} = \omega b(r)^{v} \rho$ and $P_{r} =\hat{ \omega} r^{k} R(r)^{\eta}  \rho$, respectively. Moreover, for all these particular examples, we will demonstrate that the shape functions of the corresponding exact traversable wormholes actually obey the necessary metric conditions. We also use the energy conditions in order to reveal the physical content of the constructed wormhole models. A general feature of all these models will be the violation of the NEC~($\rho + P_{i} \geq 0$) in terms of the radial pressure, $P_{r}$, at the throat of the wormhole. On the other hand, for some of the models, the violation of the DEC~($\rho - P_{i} \geq 0$) while keeping valid the NEC at the throat, turns to be possible. In other words, we will construct exact traversable wormhole models with a possible violation of the NEC and DEC at the throat of the wormhole, while $\rho \geq 0 $. That is, we have also a violation of the WEC~($\rho \geq 0$ and $\rho + P_{i} \geq 0$)~($i=r,l$ indicates radial and lateral pressure, respectively). 

In summary, we will prove in the following that, by using a varying Chaplygin gas and a varying barotropic fluid of some specific form we can reconstruct viable traversable wormhole solutions. The solutions here obtained are, to the best of our knowledge, absolutely new, and no other solutions with similar properties have ever appeared in the literature.\\

The paper is organized as follows.  In Sect.~\ref{sec:WMFE} we consider the detailed form of the field equations to be solved indicating the conditions to be satisfied by the shape function. In Sect.~\ref{sec:MEW} two exact wormhole models are obtained, to be followed by a discussion on the validity of the energy conditions. For the models considered here, we assume that the radial pressure can be described by a varying Chaplygin gas with $P_{r} = - B(r) / \rho ^{\alpha}$, where $B(r)$ can be either a function of the Ricci scalar, or a function of the shape function $b(r)$. To be able to construct exact solutions, we will concentrate our attention on the following two cases: $P_{r} = -B b(r)^{u}/ \rho^{\alpha }$ and $P_{r} = -B R(r)^{m}/ \rho^{\alpha }$. In Sect.~\ref{sec:VPF} we present two other wormhole models, corresponding to the barotropic equations of state: $P_{r} = \omega b(r)^{v} \rho$ and $P_{r} =\hat{ \omega} r^{k} R(r)^{\eta}  \rho$, respectively. Finally, Sect.~\ref{sec:Discussion} is devoted to a general discussion and conclusions.

\section{The wormhole metric and the field equations}\label{sec:WMFE}

The concept of  wormhole is one of the most popular and widely studied ones in GR and in modified theories of gravity, too~(see, e.g., \cite{Jawad:2016}~-~\cite{Zubair:2017}). Pioneering work on the static spherically symmetric wormhole, with the following metric~(in Schwarzschild coordinates $(t,r,\theta, \phi)$),
\begin{equation}\label{eq:WHMetric}
ds^{2} = -U(r) dt^{2} + \frac{dr^{2}}{V(r)} + r^{2}d\Omega^{2},
\end{equation}
where $d\Omega^{2} = d\theta^{2} + sin^{2}\theta d\phi^{2}$ and $V(r) = 1-b(r)/r$, is due to Morris and Thorne~\cite{Morris:1988}. These authors proved that the matter inside the wormhole has negative energy, thus violating the null energy condition (NEC). Later on, also dynamical wormhole models were proposed, introducing the scale factor in the metric Eq.~(\ref{eq:WHMetric}). In the literature, we meet different studies on static and dynamical wormholes, constructed using quite different hypotheses concerning the matter content. Among others, there is a line of study dedicated to speculate on possible wormholes  with the corresponding matter satisfying, e.g., the weak energy condition~(WEC) and the dominant energy condition~(DEC).  

Let us recall the parameters describing the wormhole. There is the throat, indicating the minimal surface area of the attachment. Also, in Eq.~(\ref{eq:WHMetric}), we identify the function $b(r)$, as the shape function representing the spatial shape of the wormhole. As is known, the redshift function $U(r)$ and the shape function $b(r)$ must obey the following conditions~\cite{Morris:1988}:
\begin{enumerate}
\item The radial coordinate $r$ lies between $r_{0} \leq r < \infty$, where $r_{0}$ is the throat radius.
\item At the throat, $r=r_{0}$, $b(r_{0}) = r_{0}$ and in the region outside the throat $1- b(r)/r > 0$.
\item $b^{\prime}(r_{0}) < 1$, with $\prime = d/dr$, i.e.,  the flaring out condition at the throat should be satisfied.
\item For asymptotic flatness of the space-time geometry, the limit $b(r)/r \to  0$, as $|r| \to \infty$ is required.
\item $U(r)$ must be finite and non-vanishing at the throat $r_{0}$.
\end{enumerate}
Following Refs.~\cite{Cataldo:2011} and~\cite{Rahaman:2007}, we will consider $U(r) = 1$, what means that we can achieve the de Sitter and anti-de Sitter asymptotic behaviors.  

To obtain a wormhole solution we assume also that $L_{m} = - \rho$, in order not to imply the vanishing of the extra force, and $f(\textit{R}, \textit{T}) = R + 2 f(T)$ with $f(T) =  \lambda T$~($\lambda$ is a constant). With these assumption, the field equations presented in Sect.~\ref{sec:INT} reduce to the following 
\begin{equation}\label{eq:G}
G_{ij} = (8\pi + 2\lambda) T_{ij} + \lambda (2\rho + T)g_{ij}.
\end{equation}
In Eq.~(\ref{eq:G}), $G_{ij}$ is the the usual Einstein tensor. Now, if we take into account the form of the metric, Eq.~(\ref{eq:WHMetric}), then after some algebra, for the $3$ components of the field equations, Eq.~(\ref{eq:G}), we will obtain~(see for instance~\cite{Moraes:2017c})
\begin{equation}\label{eqF1}
\frac{b^{\prime}}{r^{2}} = (8\pi + \lambda)\rho - \lambda (P_{r} + 2P_{l}),
\end{equation}
\begin{equation}\label{eqF2}
-\frac{b}{r^{3}} = \lambda \rho + (8\pi + 3\lambda)P_{r} + 2\lambda P_{l},
\end{equation}
\begin{equation}\label{eqF3}
\frac{b - b^{\prime}r}{2r^{3}} = \lambda \rho + \lambda P_{r} + (8\pi + 4 \lambda) P_{l}.
\end{equation}
To derive the above equations we have considered an anisotropic fluid satisfying the matter content of the form $T^{i}_{j} = diag(-\rho, P_{r},P_{l},P_{l})$, where $\rho = \rho (r)$ is the energy density, while $P_{r}$ and $P_{l}$ are the radial and lateral pressures, respectively. The trace $T$ of the energy-momentum tensor turns out to be $T = -\rho + P_{r} + 2P_{l}$. Moreover, it is easy to see that Eqs.~(\ref{eqF1})~-~(\ref{eqF3}) admit the following solution
\begin{equation}\label{eq:rho}
\rho = \frac{b^{\prime} }{r^{2}(8 \pi + 2 \lambda )},
\end{equation}
\begin{equation}\label{eq:Pr}
P_{r} = - \frac{b}{r^{3}(8\pi + 2\lambda )},
\end{equation}
and
\begin{equation}\label{eq:Pl}
P_{l} = \frac{b - b^{\prime}r}{2r^{3}(8\pi + 2\lambda )},
\end{equation}
for the wormhole matter content.

In the next sections, starting from two different forms for the radial pressure, we will obtain exact traversable wormhole models. We will begin our study assuming that the radial pressure can be described by a varying Chaplygn gas.

\section{Wormhole models with varying Chaplygin gas}\label{sec:MEW}

In this section, we will present wormhole models considering its matter content to be described by a varying Chaplygin gas, Eq.~(\ref{eq:VCGGEN}). Two hypothesis concerning the functional form of $B(r)$ are being put forward in this section. We will consider the shape function and the Ricci scalar parametrizations of $B(r)$. In particular $B(r) = B b(r)^{u}$ and $B(r) = B R(r)^{m}$ functional forms will be adopted during the analysis.

\subsection{Model with $P_{r} = -B b(r)^{u}/ \rho^{\alpha }$}

Let us start the search for an exact wormhole model, when the radial pressure has the following form 
\begin{equation}\label{eq:VCGCB}
P_{r} = - \frac{B b(r)^{u}}{\rho^{\alpha }},
\end{equation}
where $u$ is a constant and we keep the $b(r)$ notation to indicate that for $B(r)$ we consider a shape function parametrization. 
Now, if we take into account Eq.~(\ref{eq:Pr}) then, after integration, we can recover the functional form of the shape function by solving a first order differential equation. Therefore, the shape function of the wormhole which matter has a radial pressured given by Eq.~(\ref{eq:VCGCB}), will have the following form
\begin{equation}\label{eq:VCGCB_Br}
b(r) = \left(-\frac{(\alpha -u+1) \left(-\frac{2^{\frac{1}{\alpha }+1} (\lambda +4 \pi )^{\frac{1}{\alpha }+1} B^{1/\alpha } r^{\frac{3}{\alpha }+3}}{\frac{3}{\alpha }+3}-c_1\right)}{\alpha }\right){}^{\frac{\alpha }{\alpha -u+1}},
\end{equation}  
where $c_{1}$ is an integration constant. On the other hand, if we take into account that
\begin{equation}\label{eq:VCGCB_dBr}
b^{\prime}(r) = 2^{\frac{1}{\alpha }+1} (\lambda +4 \pi )^{\frac{1}{\alpha }+1} B^{1/\alpha } r^{\frac{3}{\alpha }+2} \hat{A}^{\frac{u-1}{\alpha -u+1}},
\end{equation}
then, after some algebra, we get
\begin{equation}\label{eq:VCGCB_rho}
\rho = 2^{1/\alpha } (\lambda +4 \pi )^{1/\alpha } B^{1/\alpha } r^{3/\alpha }\hat{A}^{\frac{u-1}{\alpha -u+1}},
\end{equation}\label{eq:VCGCB_Pl}
and 
\begin{equation}
P_{l} = \frac{\hat{A}^{\frac{\alpha }{\alpha -u+1}}-2^{\frac{1}{\alpha }+1} (\lambda +4 \pi )^{\frac{1}{\alpha }+1} B^{1/\alpha } r^{\frac{3}{\alpha }+3}\hat{A}^{\frac{u-1}{\alpha -u+1}}}{4 (\lambda +4 \pi ) r^3},
\end{equation}
where $\hat{A} = -\frac{(\alpha -u+1)}{{\alpha }} \left(-\frac{2^{\frac{1}{\alpha }+1} \alpha  (\lambda +4 \pi )^{\frac{1}{\alpha }+1} B^{1/\alpha } r^{\frac{3}{\alpha }+3}}{3 \alpha +3}-c_{1}\right)$. To obtain the last two equations we have taken into account that the energy density and pressure $P_{l}$ are given by Eq.~(\ref{eq:rho}) and Eq.~(\ref{eq:Pl}), respectively. Now, let us write also the final form of the $P_{r}$-pressure by using the form of the energy density for the matter given by Eq.~(\ref{eq:VCGCB_rho}). It is easy to see, that for $P_{r}$, we obtain 
\begin{equation}\label{eq:VCGCB_Pr}
P_{r} = -B \left(2^{1/\alpha } (\lambda +4 \pi )^{1/\alpha } B^{1/\alpha } r^{3/\alpha } \hat{A}^{\frac{u-1}{\alpha -u+1}}\right)^{-\alpha } \hat{A}^{\frac{\alpha u }{\alpha -u+1}}.
\end{equation}
A particular wormhole model with the throat at $r_{0} = 1.5$ and $b^{\prime}(1.5) \approx 0.624$ is discussed below. 
\begin{figure}[t!]
 \begin{center}$
 \begin{array}{cccc}
\includegraphics[width=80 mm]{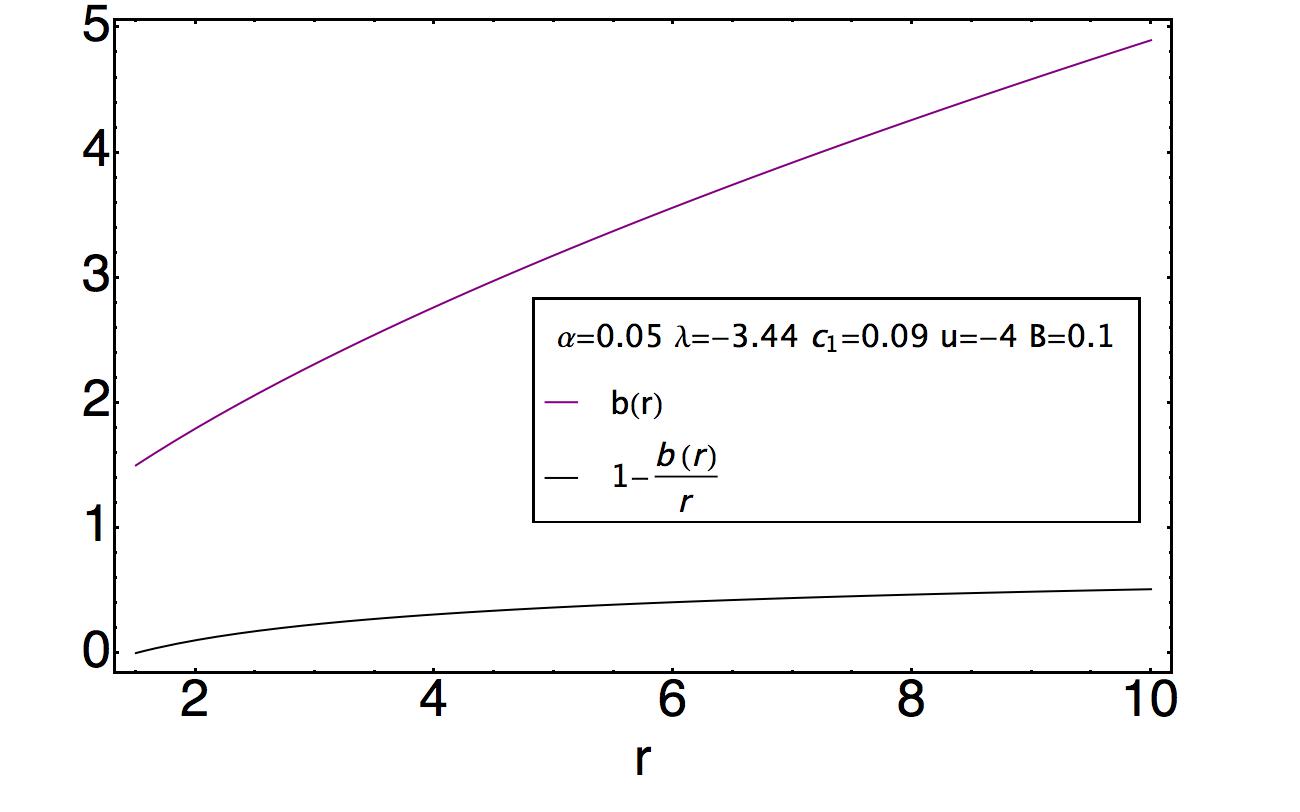} &&
\includegraphics[width=80 mm]{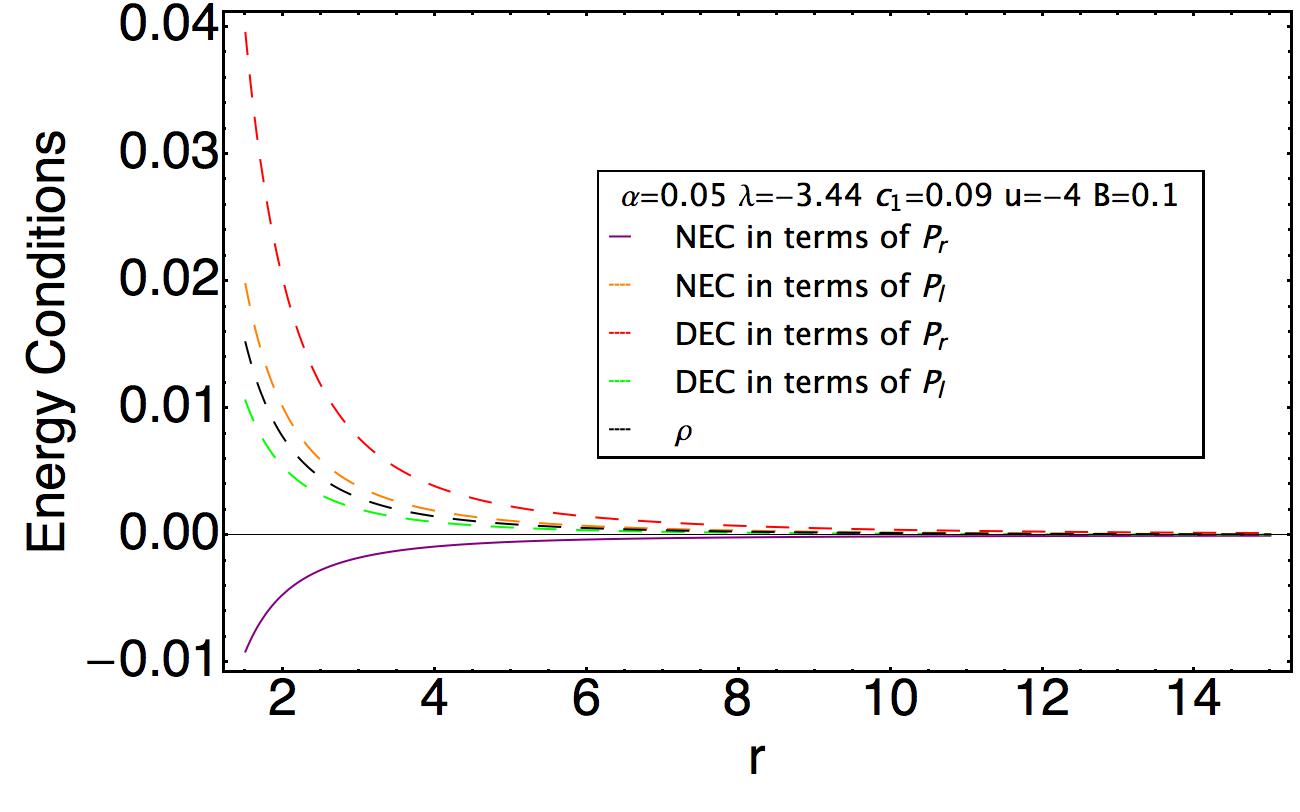}
 \end{array}$
 \end{center}
\caption{The graphical behavior of the shape function $b(r)$ for the model given by Eq.~(\ref{eq:VCGCB}) is depicted on the lhs plot. The same plot shows that the solution Eq.~(\ref{eq:VCGCB_Br}), for $b(r)$, satisfies $1- b(r)/r > 0$, for $r > r_{0}$. The throat of the wormhole occurs at $r_{0} = 1.5$, while $\alpha = 0.05$, $\lambda = -3.44$, $u = -4$, $B = 0.1$, and $c_{1} = 0.09$. The rhs plot displays the fulfillment energy conditions for the same case.}
 \label{fig:Fig1}
\end{figure}
This particular solution has been obtained for $\alpha = 0.05$, $\lambda = -3.44$, $u = -4$, $B = 0.1$, and $c_{1} = 0.09$. The graphical  behavior of the shape function $b(r)$ and $1-b(r)/r$ is presented on the left plot of Fig.~(\ref{fig:Fig1}). We can see directly that we have a solution describing a wormhole satisfying all constraints discussed in Sect.~\ref{sec:WMFE}. Moreover, the rhs plot of  Fig.~(\ref{fig:Fig1}) depicts the behavior of the energy conditions, clearly indicating that the NEC~(defined as $\rho + P_{r}$) in terms of $P_{r}$ is violated at the throat of the wormhole. We can also see that it will be valid far from the throat. On the other hand, the NEC in terms of $P_{l}$ and the DEC in terms of $P_{r}$~(defined as $\rho - P_{r}$), and the DEC in terms of $P_{l}$~(defined as $\rho - P_{l}$), are always valid. It should be mentioned that $\rho \geq 0$, what proves that only the violation of the WEC in terms of the $P_{r}$-pressure will be observed at the throat of the wormhole. The study of $\omega_{r} = P_{r}/\rho$ shows that we should expect a wormhole formation if we have a phantom varying Chaplygin gas as given by Eq.~(\ref{eq:VCGCB}).    

\begin{figure}[t!]
 \begin{center}
$ \begin{array}{cccc}
\includegraphics[width=80 mm]{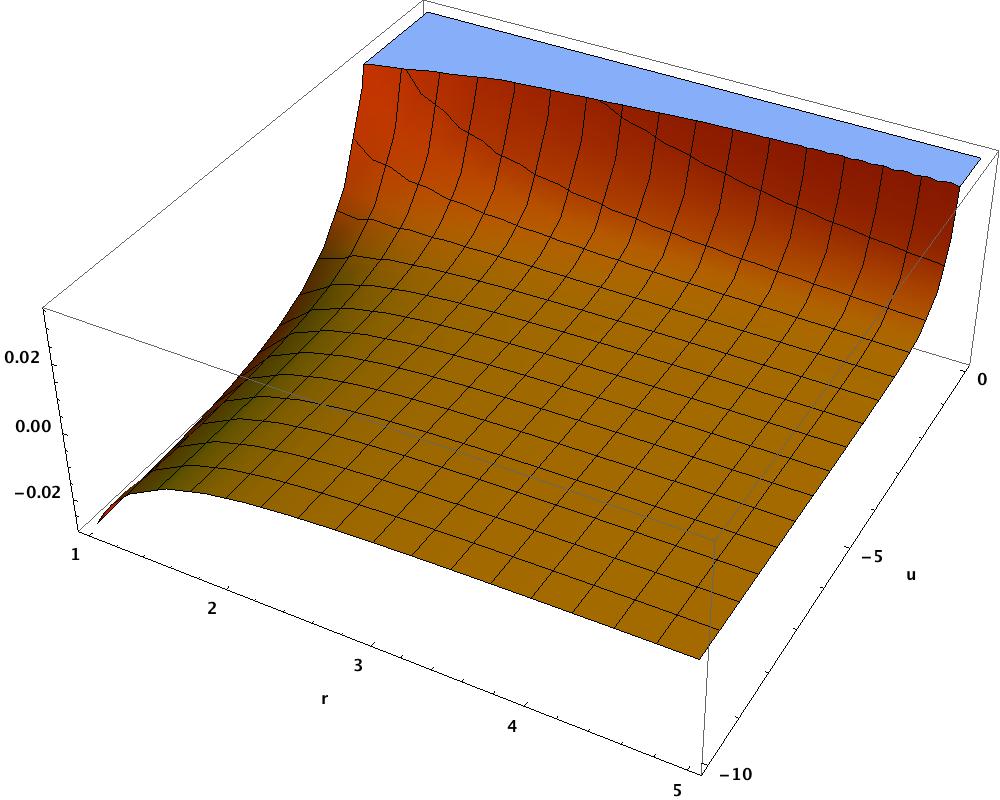} &&
\includegraphics[width=80 mm]{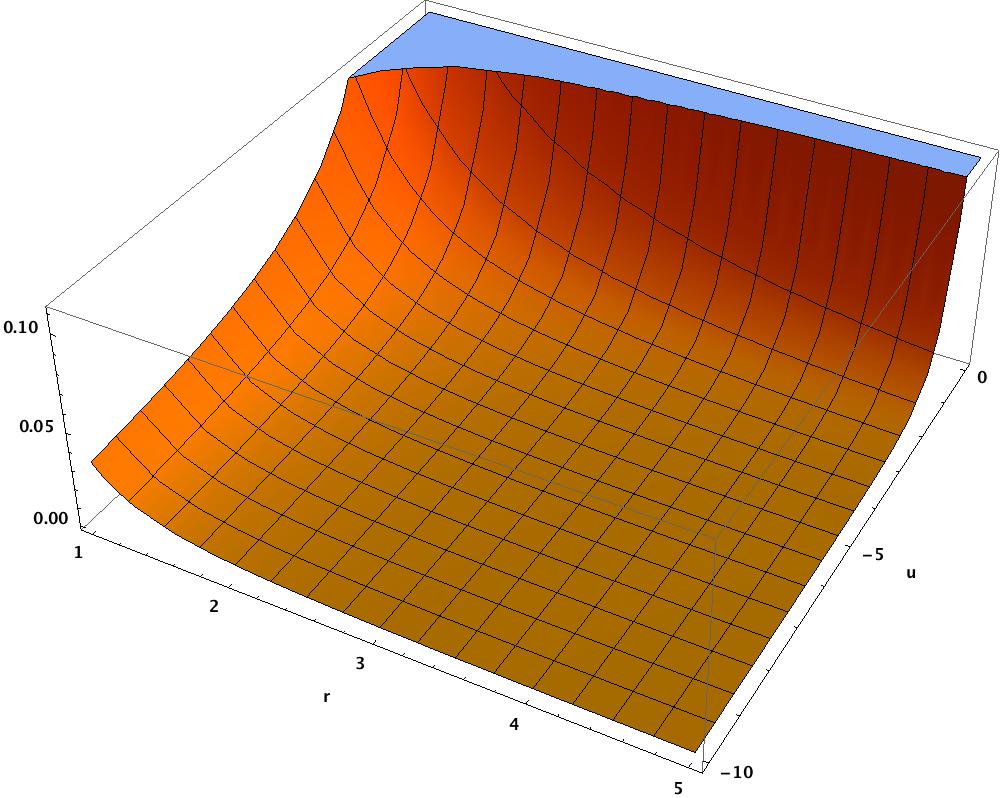}\\
\includegraphics[width=80 mm]{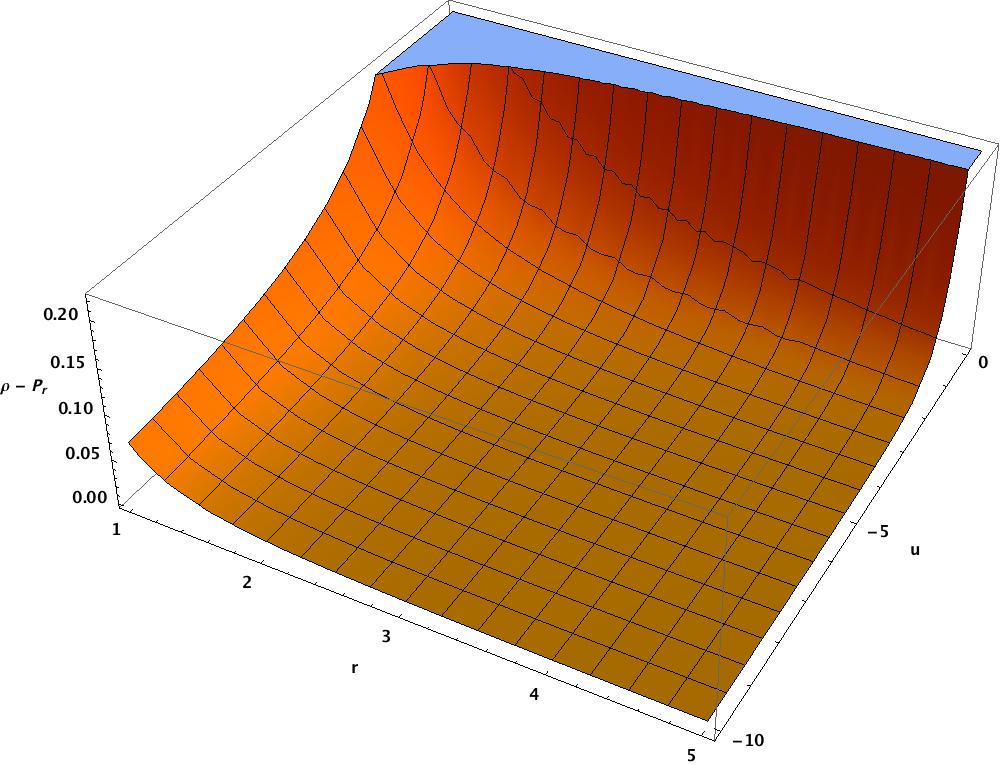} &&
\includegraphics[width=80 mm]{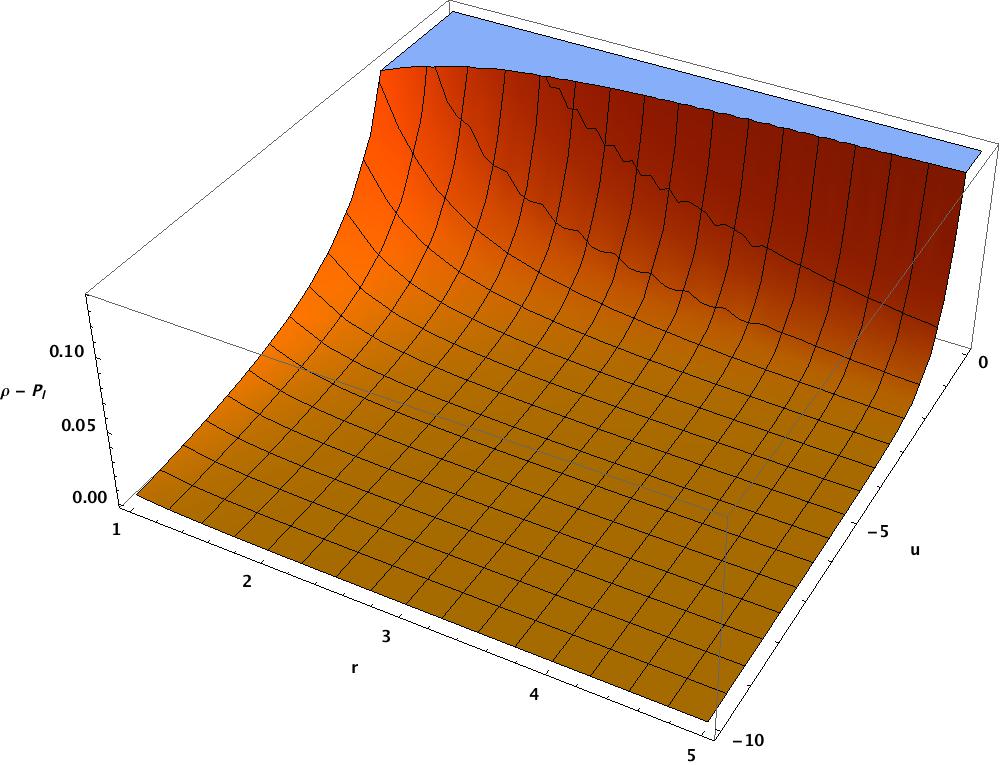}
 \end{array} $
 \end{center}
\caption{The graphical behavior of the NEC in terms of the $P_r$ and $P_l$ pressures is depicted on the two Plots\_EPS of the top panel. We see the existence of regions where the NEC is fulfilled, in terms of both pressures. The bottom panel shows the graphical behavior of the DEC in terms of both pressures. We see that the DEC in terms of both pressures can be valid even for the region where the NEC in terms of $P_r$ is not valid. The shape function $b(r)$ for the model is given by Eq.~(\ref{eq:VCGCB_Br}). The plot has been obtained for $\alpha = 0.5$, $\lambda = 5$, $B = 1$, and $c_1 = 1$, and for different values of $u$.}
 \label{fig:Fig1_1}
\end{figure}

In Fig.~(\ref{fig:Fig1_1}), we present the graphical behavior of the NEC and the DEC in terms of the $P_{r}$ and $P_{l}$ pressures, for different values of the parameter $u$, when $\alpha = 0.5$, $\lambda = 5$, $b = 1.0$, and $c_{1} = 1.0$. The top panel in Fig.~(\ref{fig:Fig1_1}) corresponds to the behavior of NEC in terms of the $P_{r}$ and $P_{l}$ pressures. We clearly see that there are regions where the NEC in terms of $P_{r}$ is not valid, while the NEC in terms of $P_{l}$ is valid. Moreover, we see that, for $u > -2$, we can have regions where the NEC in terms of $P_{r}$ is also valid. On the other hand, the bottom panel shows the graphical behavior of the DEC in terms of the two pressures. We observe that even for the regions where the NEC in terms of $P_{r}$ is violated, the DEC in terms of both pressures continues to be valid. Moreover, for the regions and range of  the parameters considered, we observe that $\rho \geq 0$ is satisfied.

\subsection{Model with $P_{r} = -B R(r)^{m}/ \rho^{\alpha }$}

Here we turn our attention to the model with
\begin{equation}\label{eq:VCGR}
P_{r} = -\frac{B R(r)^{m}}{ \rho^{\alpha }},
\end{equation}
where $R(r) = 2 b^{\prime}/r^{2}$ is the Ricci scalar. It is easy to see, that similarly to the model given by Eq.~(\ref{eq:VCGCB}), we can also obtain exact wormhole solutions. A particular one can be found if we consider  $m=\alpha $. For instance, if we take $m=\alpha =1$, then the shape function  will admit the following form
\begin{equation}
b(r) = 8 B (\lambda +4 \pi )^2 r^3.
\end{equation}
On the other hand, if we consider $m=\alpha = 4$, then for the shape function we get
\begin{equation}
b(r) = 512 B (\lambda +4 \pi )^5 r^3.
\end{equation}
Now, if we consider $-m=\alpha = 1$, then direct integration of Eq.~(\ref{eq:Pr}) will yield two solutions, for the shape function $b(r)$, as 
\begin{equation}
b_{1,2}(r) = \frac{\left(\mp 2 \sqrt{2} \sqrt{B} \lambda  r^{9/2}-8 \sqrt{2} \pi  \sqrt{B} r^{9/2}+9 c_2\right){}^{2/3}}{6^{2/3}},
\end{equation}
where $c_{2}$ is an integration constant. Eventually, another exact wormhole solution is found, given by
\begin{equation}\label{eq:NEC3D}
b(r) = c_3 r^{\frac{1}{16 B (\lambda +4 \pi )^2}},
\end{equation}
with $c_{3}$ an integration constant, when we consider the parameters $m$ and $\alpha$ to be $m=2$ and $\alpha=1$, respectively. Then, each form of the shape function considered above can be used to obtain the explicit form of $\rho$, $P_{r}$, and $P_{l}$. More general case can be studied numerically, which is  done below for the wormhole solution with $B = 0.001$, $\alpha = 0.95$, $\lambda = 9.5$ and $m=1.95$. 
\begin{figure}[t!]
 \begin{center}$
 \begin{array}{cccc}
\includegraphics[width=80 mm]{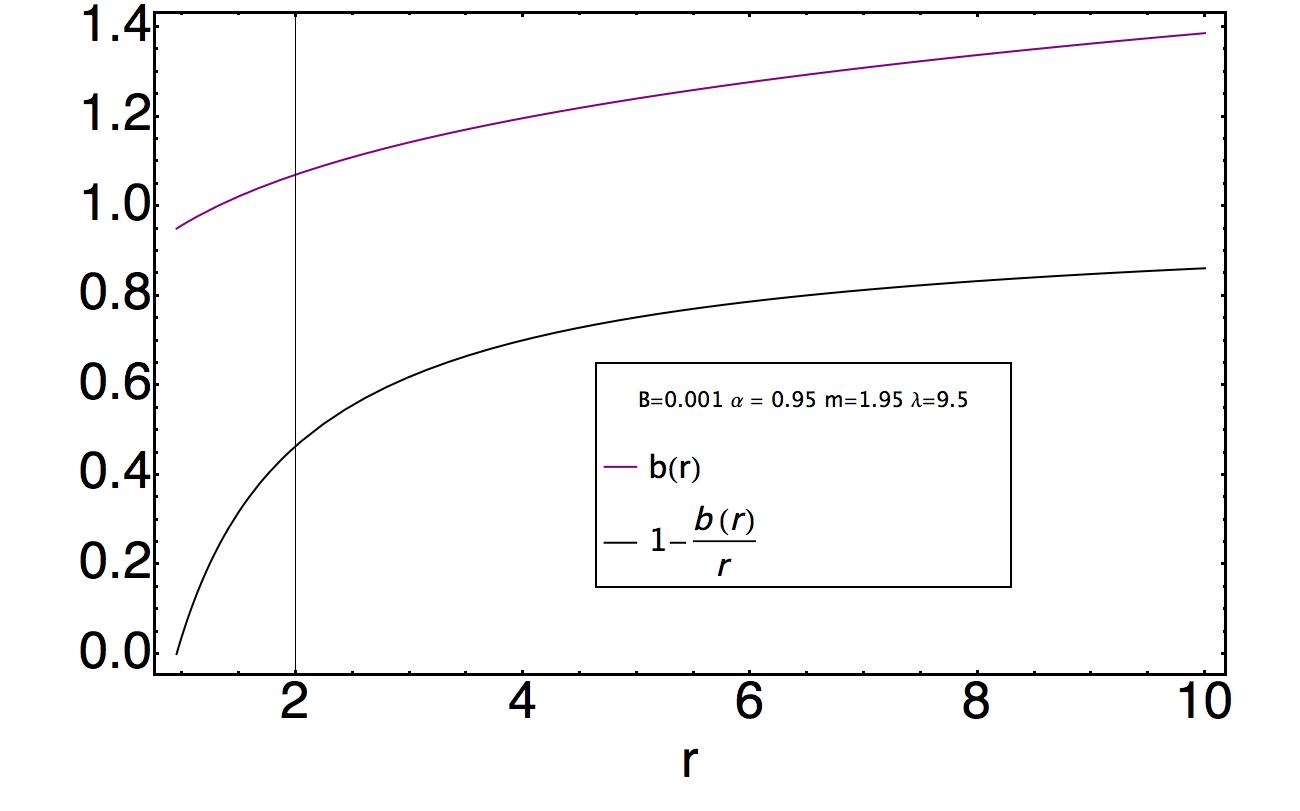} &&
\includegraphics[width=80 mm]{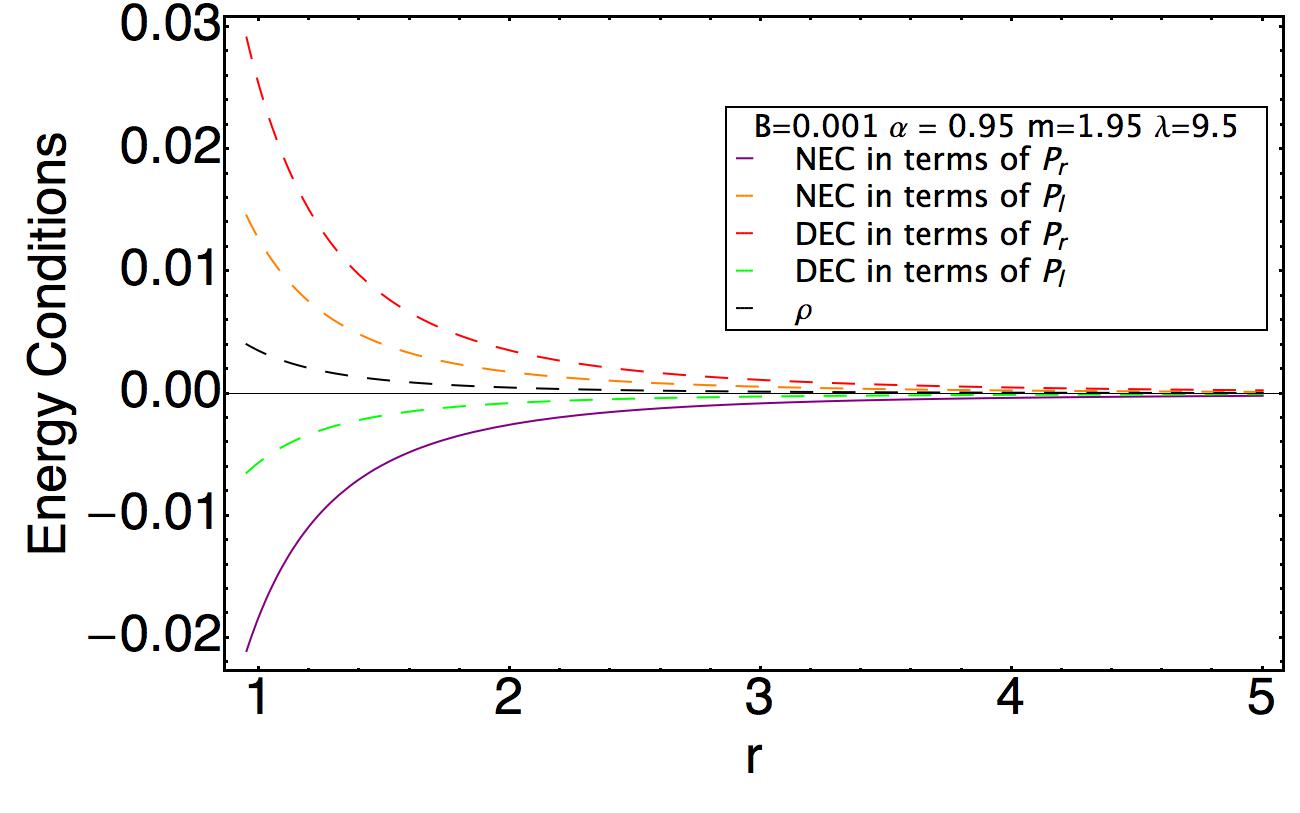}
 \end{array}$
 \end{center}
\caption{The graphical behavior of the shape function $b(r)$ for the model given by Eq.~(\ref{eq:VCGR}) is presented on the lhs plot. We observe that the solution for $b(r)$ satisfies $1- b(r)/r > 0$, for $r > r_{0}$. The throat of the wormhole occurs at $r_{0} = 0.95$, while $\alpha = 0.95$, $\lambda = 9.5$, $m = 1.95$, and $B = 0.001$. The rhs plot proves the fulfillment of the energy conditions for the same case.}
 \label{fig:Fig2}
\end{figure}

The throat of this wormhole is obtained to occur for $r_{0} = 0.95$, while $b^{\prime}(0.95) \approx 0.161$. From the graphical behavior of the NEC in terms of $P_{r}$, presented in Fig.~(\ref{fig:Fig2}), we see that it will be violated at the throat of such wormhole, while it will be valid far from the throat. Moreover, we should expect the violation of the DEC in terms of $P_{l}$ at the throat, while, similar to the NEC in terms of $P_{r}$, it will be valid far from the throat. On the other hand, we should expect to have a valid NEC in terms of $P_{l}$ and a valid DEC in terms of the $P_{r}$ pressure at the throat of the wormhole. They are also valid far from the throat. We also have $\rho > 0$ at the throat and $\rho = 0$ far from the throat; which means that violation of the WEC in terms of $P_{r}$ should be expected as well. On the other hand, the WEC in terms of $P_{l}$ will be valid at the throat and also far from it. The lhs plot of Fig.~(\ref{fig:Fig2}) clearly shows that we have a wormhole solution. Furthermore, the study of the model with $P_{r} = A\rho - B R(r)^{m}/ \rho^{\alpha }$, shows that $A$ should be of the same order as $B$ in order to yield a traversable wormhole solution. The numerical study shows that in this case we will obtain the energy conditions having qualitatively the same behavior as in case of $A = 0$ presented above.       

\begin{figure}[t!]
 \begin{center}$
 \begin{array}{cccc}
\includegraphics[width=80 mm]{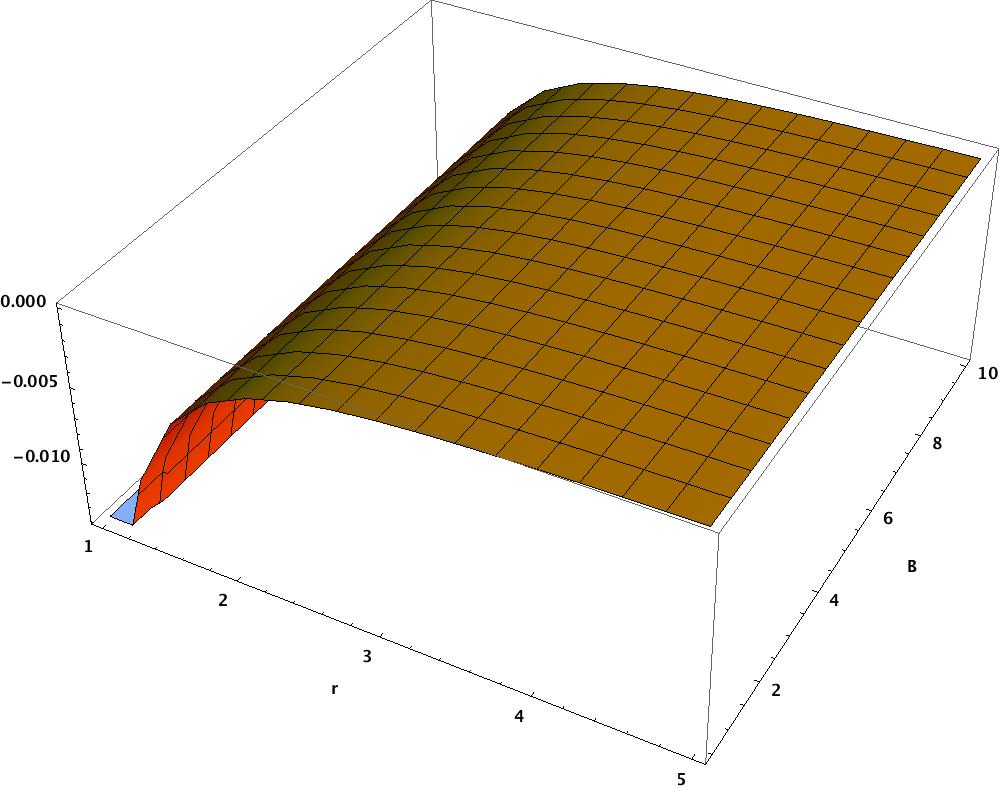} &&
\includegraphics[width=80 mm]{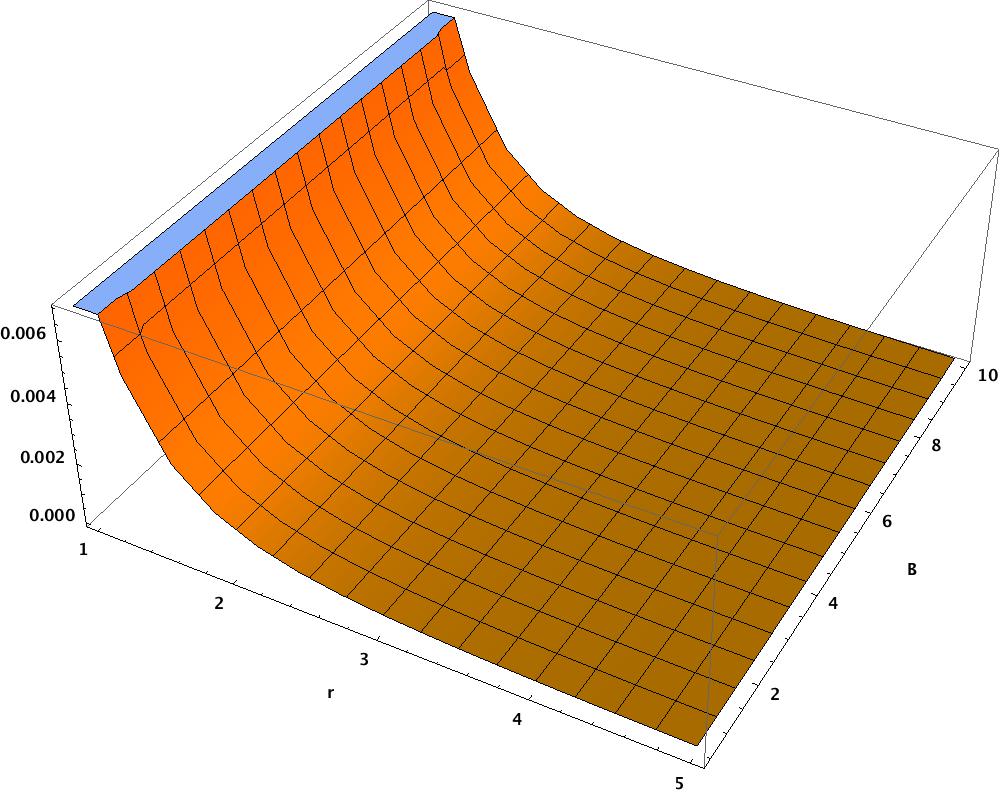}\\
 \end{array}$
 \end{center}
\caption{The graphical behavior of NEC in terms of $P_{r}$ is presented on the lhs plot. The rhs plot corresponds to the graphical behavior of the NEC in terms of the $P_{l}$ pressure. In both cases, we have considered the validity of the NEC for different values of the parameter $B$. The shape function $b(r)$ for this case is given by Eq.~(\ref{eq:NEC3D}).}
 \label{fig:Fig2_1}
\end{figure}

In Fig.~(\ref{fig:Fig2_1}) we depict the graphical behavior of NEC in terms of both pressures, for different values of the parameter $B$. In particular, we observe that for fixed values of the parameters $m$ and $\alpha$, the parameters $B$ and $\lambda$ cannot affect the behavior of the energy conditions. From Fig.~(\ref{fig:Fig2_1}) we see that, for a small value of $r$, the NEC in terms of the  $P_{r}$ pressure is always violated, while the same energy condition in terms of the $P_{l}$ pressure is valid. On the other hand, the study shows also, that DEC in terms of $P_{l}$ is also not valid for the regions with small $r$.
  
\section{Wormhole models with varying barotropic fluid}\label{sec:VPF}

In this section we will consider other exact wormhole models obtained from the assumption that the radial pressure of the matter content of the wormhole can be described by a varying barotropic fluid
\begin{equation}
P_{r} = \omega(r) \rho,
\end{equation} 
where $\omega(r)$ is the varying equation of state parameter. In particular, similarly to the varying Chaplygin gas models considered above, we will assume the shape function and the Ricci scalar parametrizations of the equation of state parameter.

\subsection{Model with $P_{r} = \omega b(r)^{v} \rho$}

\begin{figure}[t!]
 \begin{center}$
 \begin{array}{cccc}
\includegraphics[width=80 mm]{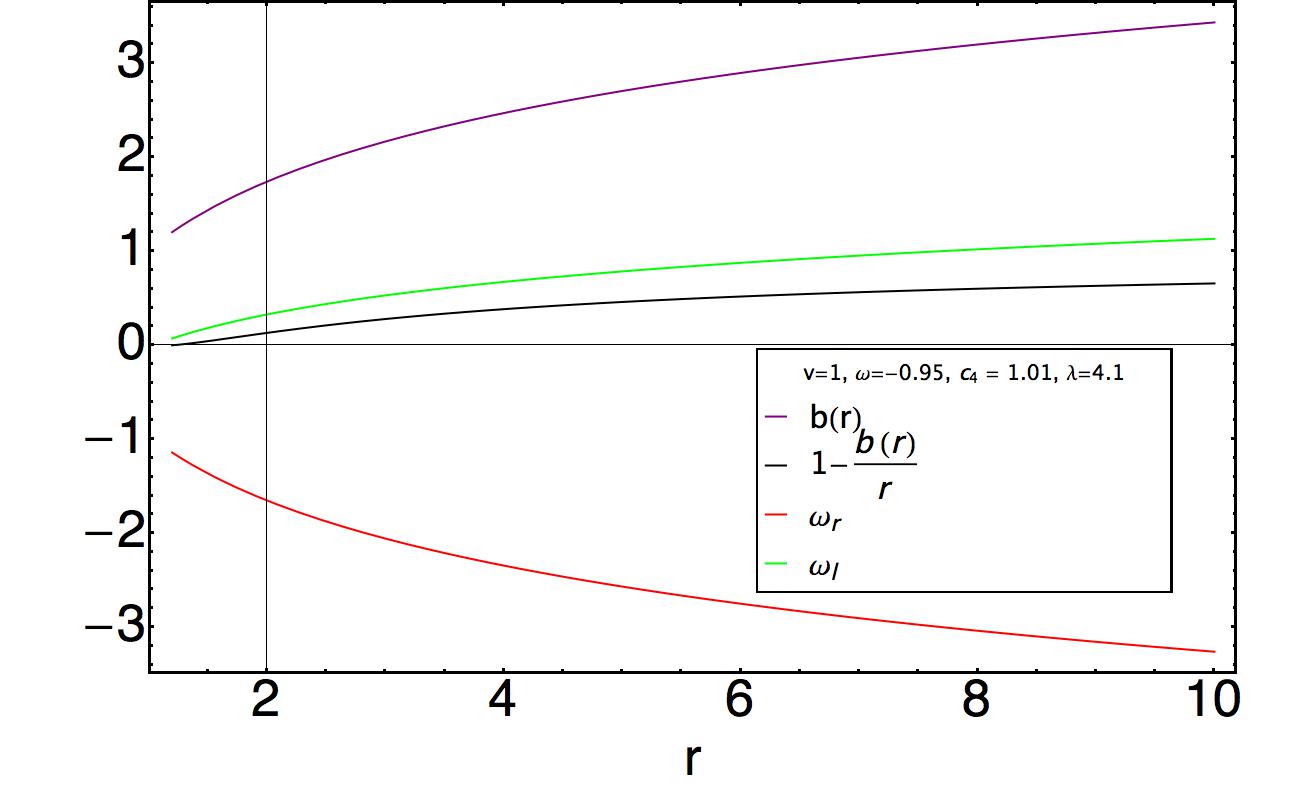} &&
\includegraphics[width=80 mm]{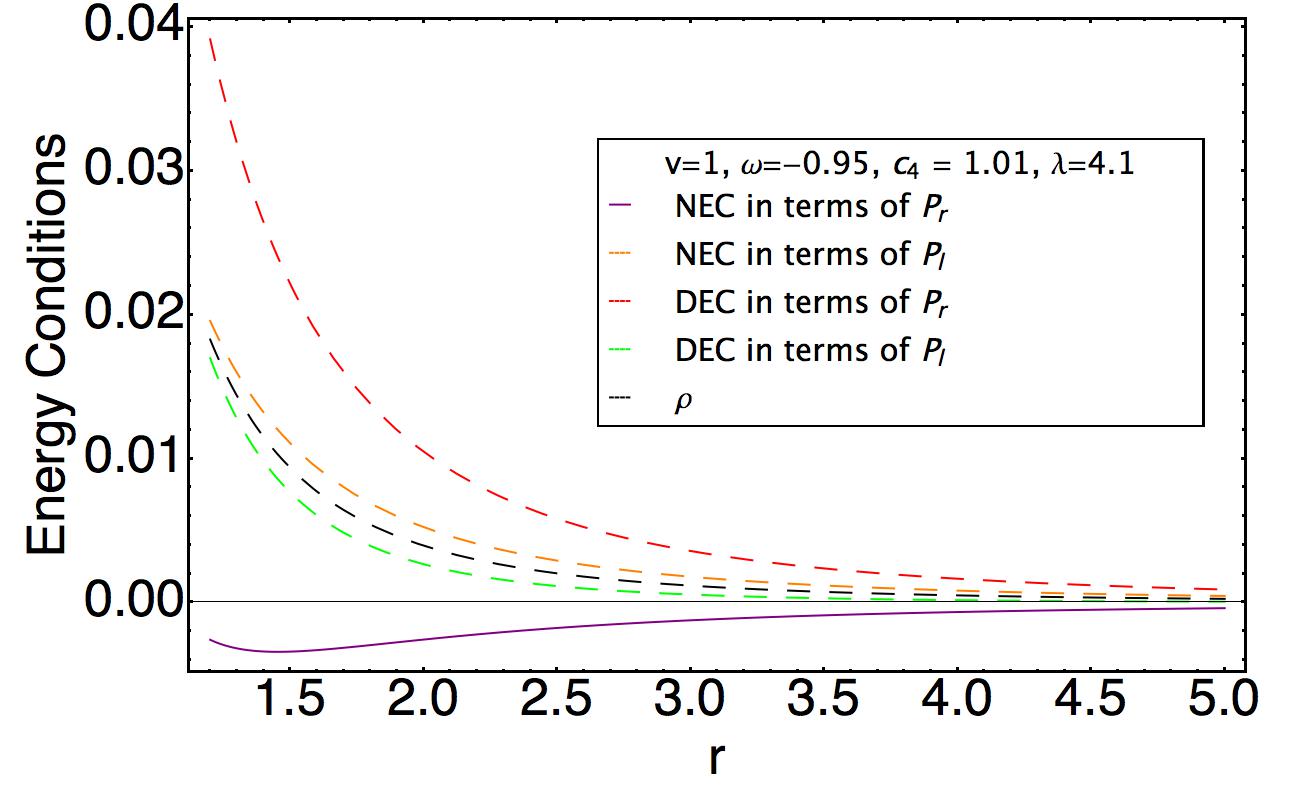}
 \end{array}$
 \end{center}
\caption{The graphical behavior of the shape function $b(r)$ for the model given by Eq.~(\ref{eq:GVPG_Pr1}) is presented on the lhs plot. We wee from there  that the solution, Eq.~(\ref{eq:VPG_br1}), for $b(r)$ satisfies $1- b(r)/r > 0$, for $r > r_{0}$. The  behaviors of $\omega_{r} = P_{r}/\rho$ and $\omega_{l} = P_{l}/\rho$ are given in the same plot. The throat of the wormhole occurs for $r_{0} = 1.2$, while $v=1$, $\omega = -0.95$, $c_{4}=1.01$, and $\lambda = 4.1$. The rhs plot shows the fulfillment of the energy conditions for the same case.}
 \label{fig:Fig3}
\end{figure}

The first wormhole model of this section has been obtained assuming that 
\begin{equation}\label{eq:GVPG_Pr1}
P_{r} = \omega  b(r)^{v} \rho.
\end{equation} 
This means that we consider a varying equation of state parametrization according to the shape function $b(r)$. On the other hand, it is easy to see that the following form of the shape function describing the wormhole
\begin{equation}\label{eq:VPG_br1}
b(r) = \left(v \left(c_{4}-\frac{\log (r)}{\omega }\right)\right){}^{1/v},
\end{equation} 
where $c_{4}$ is an integration constant, will be obtained from the integration of Eq.~(\ref{eq:Pr}). If the form of the shape function is known then, from Eqs.~(\ref{eq:rho}), (\ref{eq:Pr}), and (\ref{eq:Pl}), we get
\begin{equation}\label{eq:VPG_rho1}
\rho = -\frac{\left(v \left(c_4-\frac{\log (r)}{\omega }\right)\right){}^{\frac{1}{v}-1}}{2 (\lambda +4 \pi ) r^3 \omega },
\end{equation}
\begin{equation}\label{eq:VPG_Pr1}
P_{r} = -\frac{\omega  \left(\left(v \left(c_4-\frac{\log (r)}{\omega }\right)\right){}^{1/v}\right){}^{v+1}}{2 (\lambda +4 \pi ) r^3 v \left(c_4 \omega -\log (r)\right)},
\end{equation}
and
\begin{equation}\label{eq:VPG_Pl1}
P_{l} = \frac{\left(c_4 v \omega -v \log (r)+1\right) \left(v \left(c_4-\frac{\log (r)}{\omega }\right)\right){}^{1/v}}{4 (\lambda +4 \pi ) r^3 v \left(c_4 \omega -\log (r)\right)}.
\end{equation}
A particular wormhole model described by Eq.~(\ref{eq:VPG_br1}), Eqs.~(\ref{eq:VPG_rho1}),~(\ref{eq:VPG_Pr1}), and~(\ref{eq:VPG_Pl1}) is obtained for $v=1$, $\omega = -0.95$, $c_{4}=1.01$, and $\lambda = 4.1$. The behavior of the shape function $b(r)$ and of $1-b(r)/r$ is depicted on the lhs  of Fig.~(\ref{fig:Fig3}). 

Moreover, the same Plots\_EPS depict the behaviors of the radial and lateral equation of state parameters, indicating that wormhole formation is possible provided the radial equation of state parameter has a phantom nature. On the other hand, and at the same time, the lateral equation of state parameter should have a behavior leading to $\omega_{l} > 0$. The throat of this particular wormhole occurs at $r_{0} = 1.2$ and $b^{\prime}(r_{0}) \approx 0.877$.  Furthermore, from the right plot in Fig.~(\ref{fig:Fig3}), we see that only the NEC in terms of the radial pressure $P_{r}$ will be violated at the throat of the wormhole. This implies also that the WEC in terms of $P_{r}$ will be violated at the throat, since there $\rho > 0$. We can confirm also that the NEC and the WEC in terms of $P_{r}$  will eventually become valid for large values of $r$. We see also, that the DEC in terms of both pressures is valid at the throat of the wormhole, too. Together with the plot of the shape function $b(r)$ we depict the behavior of the radial $\omega_{r} = P_{r}/\rho$ and lateral $\omega_{l} = P_{l}/\rho$ equation of state parameters, as well. The behavior of $\omega_{r}$ indicates that it has a phantom nature and, thus, wormhole formation is possible provided the lateral equation of state parameter satisfies $\omega_{l} > 0$.

\begin{figure}[t!]
 \begin{center}$
 \begin{array}{cccc}
\includegraphics[width=80 mm]{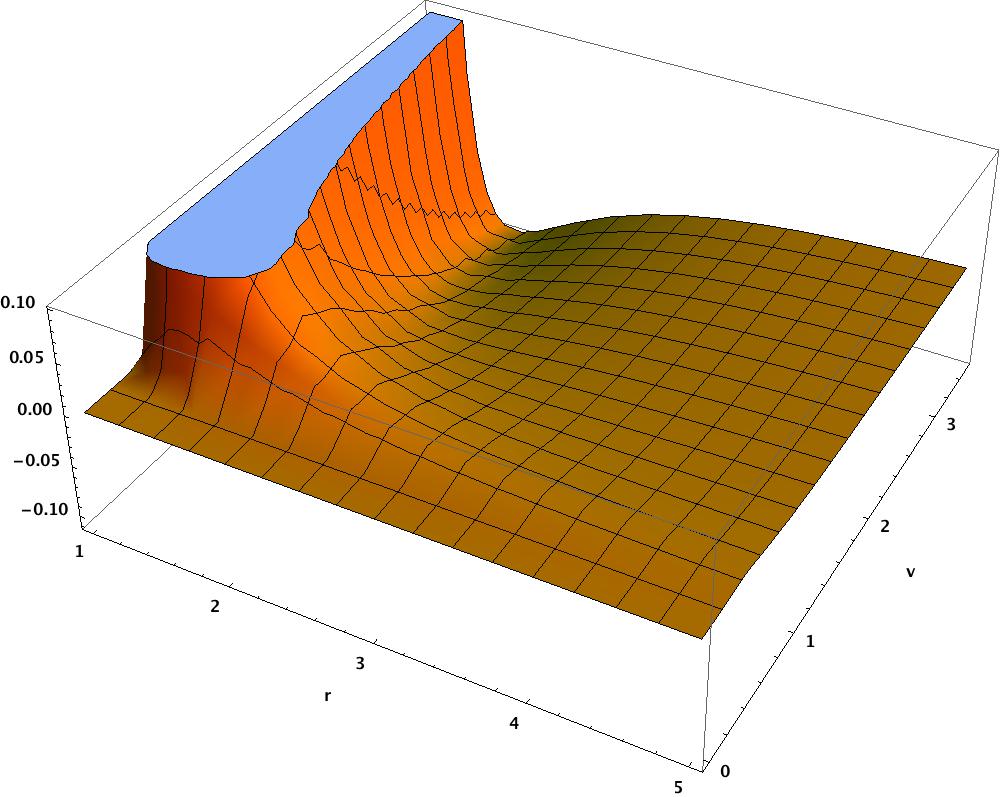} &&
\includegraphics[width=80 mm]{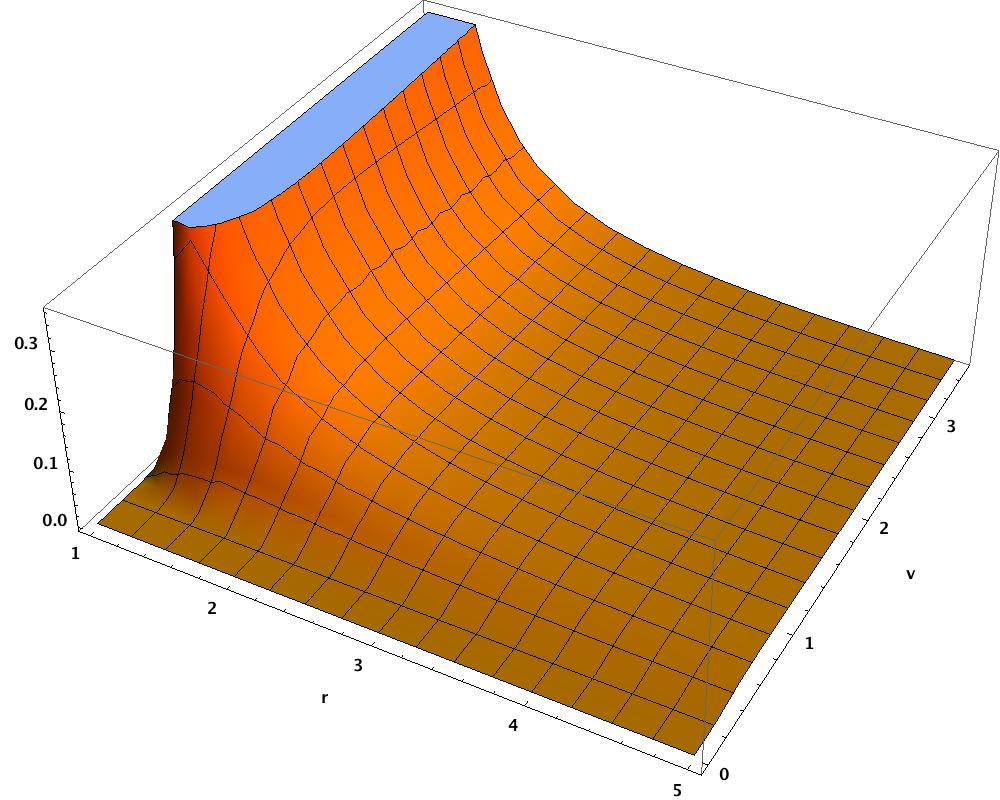}\\
\includegraphics[width=80 mm]{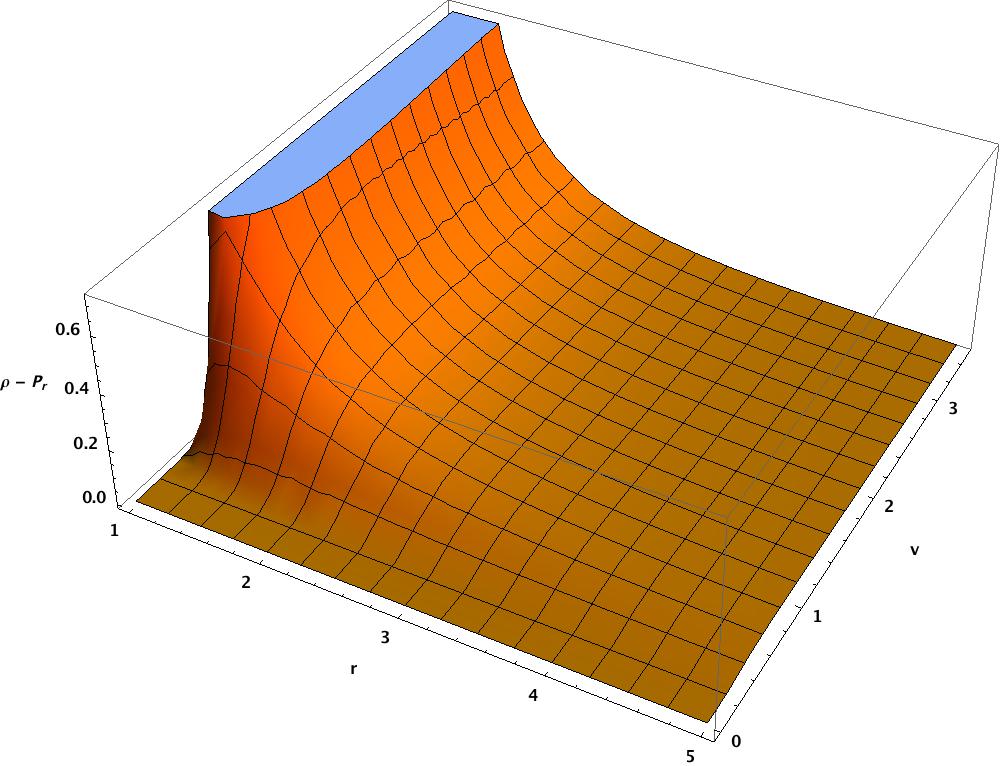} &&
\includegraphics[width=80 mm]{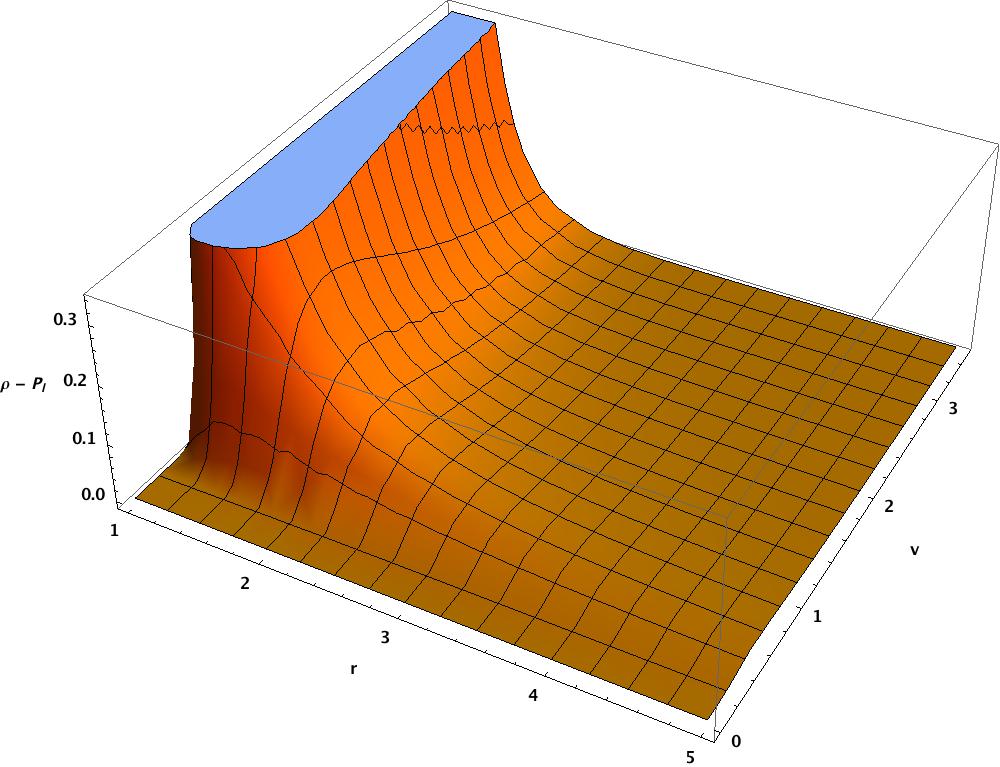}
 \end{array}$
 \end{center}
\caption{The graphical behavior of the NEC in terms of both pressures for different values of $v$ parameter is presented on the top panel. The bottom panel depicts the behavior of the DEC in terms of the pressures $P_{r}$ and $P_{l}$,  for different values of $v$ parameter. The shape function $b(r)$ for the model is given by Eq.~(\ref{eq:VPG_br1}). It clearly shows that the NEC in terms of $P_{r}$ for small $r$ and big values of $v$ can be violated, while the DEC in terms of both pressures and the NEC in terms of $P_{r}$ remain valid. The regions where both energy conditions are valid are also shown.}
 \label{fig:Fig3_1}
\end{figure}

To conclude this subsection, we would like to discuss the impact of the parameter $v$ on the energy conditions. In particular, the graphical distribution of the validity regions of the NEC and DEC, in terms of both pressures for $v \in [0,3.5]$ is presented in Fig.~(\ref{fig:Fig3_1}). The top panel corresponds to the behavior of the NEC in terms of the pressures  $P_{r}$ and $P_{l}$. From the behavior of the NEC in terms of $P_{r}$ depicted on the lhs plot, we see that a local violation of the same, for higher values of $v$ parameter, is possible, while the NEC in terms of $P_{l}$ will still be valid. On the other hand, from the bottom panel of the same figure we conclude that the validity of the DEC in terms of both pressures will be maintained. In other words, the parameter $v$ can have a non-trivial impact on the fulfillment of the NEC. We would like to mention, that the condition $\rho \geq 0$ is valid even when NEC is violated. The regions and appropriate values for the NEC and DEC in terms of both pressures to be valid can be estimated from the plots of Fig.~(\ref{fig:Fig3_1}). 

In the next subsection we will consider a different wormhole model, by assuming the following varying barotropic equation of state $P_{r} =\hat{ \omega} r^{k} R(r)^{\eta}  \rho$ for the radial pressure. Actually, the consideration of this specific form for the radial pressure is the final result of the study started with the assumption of a radial pressure of the generic type $P_{r} =\hat{ \omega} R(r)^{\eta}  \rho$. Our study concludes that although we can obtain exact solutions for the shape function, however, the solutions will describe non-traversable wormhole models. Anyway, it is well known that in such cases, in theory, we could just glue an exterior flat geometry into the interior geometry and solve the traversability problem. On the other hand, the form for the radial pressure to be considered in the next section will allow us to overcome this problem, i.e. we will indeed obtain solutions describing traversable wormholes.

\subsection{Model with $P_{r} =\hat{ \omega} r^{k} R(r)^{\eta}  \rho$}

Let us here consider another family of wormhole models, derived from the assumption that
\begin{equation}\label{eq:GVPG_Pr2}
P_{r} = \hat{\omega} r^{k} R(r)^{\eta} \rho,
\end{equation} 
where $\hat{\omega}$, $k$, and $\eta$ are constant, while $R(r)$ is the Ricci scalar. To simplify the discussion, we will consider particular values for the $k$ and $\eta$ parameters. It is easy to see, that in case of $\eta = -2$ and $k=-5$, one obtains an exact wormhole solution, described by the following shape function
\begin{equation}\label{eq:VPG_br2}
b(r) = \sqrt{2 c_5 - \frac{r \hat{\omega} }{2}},
\end{equation}
where $c_{5}$ is an integration constant. Then, after some algebra, we  get
\begin{equation}\label{eq:VPG_rho2}
\rho =-\frac{\hat{\omega} }{4 (\lambda +4 \pi ) r^2 \sqrt{8 c_{5}-2 r \hat{\omega} }},
\end{equation}
\begin{equation}\label{eq:VPG_Pr2}
P_{r} = -\frac{\sqrt{8 c_{5}-2 r \hat{\omega} }}{4 (\lambda +4 \pi ) r^3},
\end{equation}
and
\begin{equation}\label{eq:VPG_Pl2}
P_{l} = \frac{8 c_{5}-r \hat{\omega} }{8 (\lambda +4 \pi ) r^3 \sqrt{8 c_{5}-2 r \hat{\omega} }}.
\end{equation}

\begin{figure}[t!]
 \begin{center}$
 \begin{array}{cccc}
\includegraphics[width=80 mm]{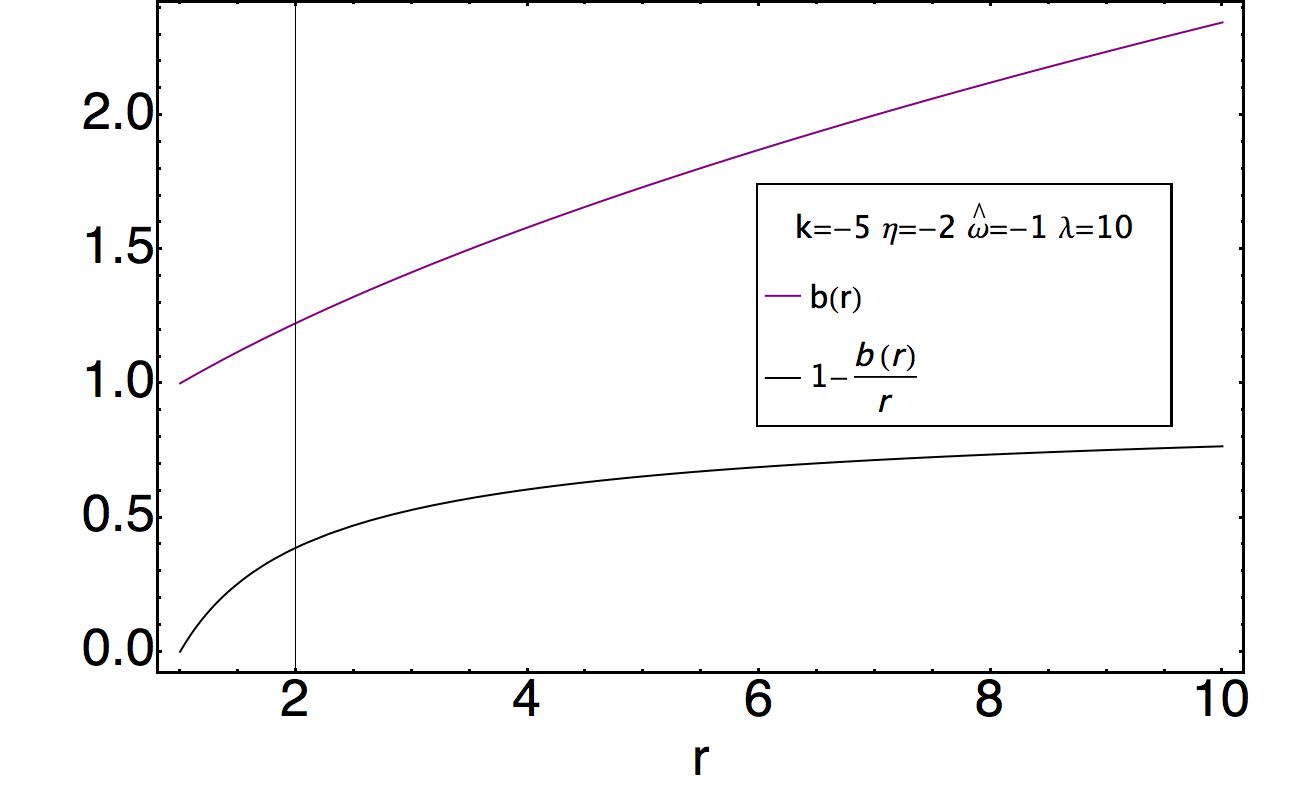} &&
\includegraphics[width=80 mm]{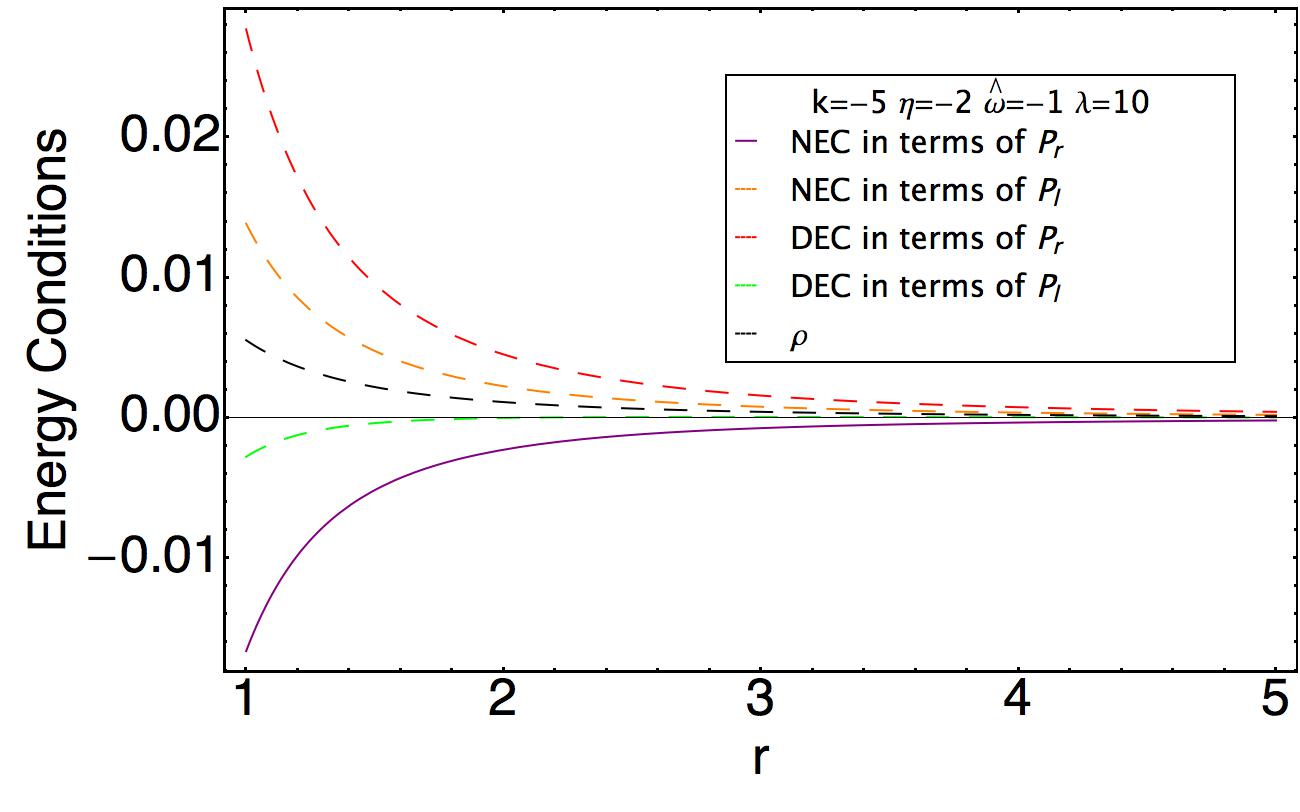}
 \end{array}$
 \end{center}
\caption{The graphical behavior of the shape function $b(r)$ for the model given by Eq.~(\ref{eq:GVPG_Pr2}) is presented on the lhs plot. The same plot shows that the solution Eq.~(\ref{eq:VPG_br2}) for $b(r)$ satisfies $1- b(r)/r > 0$, for $r > r_{0}$. The throat of the wormhole occurs at $r_{0} = 1.0$, while $k = -5$, $\eta = -2$, $\lambda = 10$, and $\hat{\omega} = -1$. The rhs plot corresponds to the energy conditions for the same case obtained from Eqs.~(\ref{eq:VPG_NECPr2}),~(\ref{eq:VPG_NECPl2}),~(\ref{eq:VPG_DECPr2}), and (\ref{eq:VPG_DECPl2}), respectively.}
 \label{fig:Fig4}
\end{figure}

Now, let us consider
\begin{equation}\label{eq:VPG_NECPr2}
\rho + P_{r} = \frac{r \hat{\omega} -8 c_{5}}{4 (\lambda +4 \pi ) r^3 \sqrt{8 c_{5}-2 r \hat{\omega} }},
\end{equation}
\begin{equation}\label{eq:VPG_NECPl2}
\rho + P_{l} = \frac{8 c_{5}-3 r \hat{\omega} }{8 (\lambda +4 \pi ) r^3 \sqrt{8 c_{5}-2 r \hat{\omega} }} ,
\end{equation}
\begin{equation}\label{eq:VPG_DECPr2}
\rho - P_{r} = \frac{8 c_{5}-3 r \hat{\omega} }{4 (\lambda +4 \pi ) r^3 \sqrt{8 c_{5}-2 r \hat{\omega} }},
\end{equation}
and
\begin{equation}\label{eq:VPG_DECPl2}
\rho - P_{l} = \frac{-8 c_{5}-r \hat{\omega} }{8 (\lambda +4 \pi ) r^3 \sqrt{8 c_{5}-2 r \hat{\omega} }},
\end{equation}
the explicit forms of the NEC and DEC in terms of the pressures $P_{r}$ and $P_{l}$, respectively. The graphical behavior of the shape function $b(r)$, Eq.~(\ref{eq:VPG_br2}), and of $1-b(r)/r$, is depicted on the lhs plot of Fig.~(\ref{fig:Fig4}), for $\hat{\omega} = -1$ and $\lambda = 10$. Moreover, the throat of this model occurs for $r_{0} = 1$ and $b^{\prime}(r_{0}) = 0.25$. This means that we have indeed obtained a realistic exact wormhole model. At the end of this subsection we will also report the results from the study of the energy conditions. In particular, from the graphical behavior of the NEC, DEC, and $\rho$ available on the rhs plot of Fig.~(\ref{fig:Fig4}), we can claim that the NEC and WEC in terms of $P_{r}$ are not fulfilled at the throat of this wormhole. On the other hand, the DEC in terms of $P_{l}$ is also not fulfilled at the throat. Similarly to the other three models above, the validity of all energy conditions in terms of both pressures will be observed for large values of $r$, that is, far from the throat of the wormhole. We should also mention that the NEC and the WEC in terms of $P_{l}$, with the DEC in terms of $P_{r}$, are valid everywhere.     

\section{\large{Discussion and conclusions}}\label{sec:Discussion}

Dark energy was introduced in order to address the issue of the accelerated expansion of the Universe, a problem that can be much alleviated if we consider modifications of GR. Modified theories of gravity can very efficiently work also for the dark matter problem. Basically, there are two different ways to modify GR, namely we can  either modify the geometrical or the matter content of the  action of this theory. In the present paper we have considered a very specific modification of GR known as a   
$f(\textit{R}, \textit{T})$ theory of gravity, where $T = \rho + P_{r} + 2P_{l}$ is the trace of the energy-momentum tensor. We have studied the wormhole model formation problem in the case of a rather specific form of $f(\textit{R}, \textit{T})$, for four different types of assumptions concerning  the description of the corresponding radial pressure. In particular, we have considered the following modification: $f(\textit{R}, \textit{T}) = R + 2 \lambda T$, assuming that the radial pressure can be described either by the equation of state of a varying Chaplygin gas, or by the equation of state parameter of a varying barotropic fluid. 

In the first part of the paper, we have considered the following two forms for the varying Chaplygin gas: $P_{r} = -B b(r)^{u}/ \rho^{\alpha }$ and $P_{r} = -B R(r)^{m}/ \rho^{\alpha }$, respectively. In the second part, we have used the following two forms for the varying barotropic fluid: $P_{r} = \omega b(r)^{v} \rho$ and  $P_{r} =\hat{ \omega} r^{k} R(r)^{\eta}  \rho$, in order  to describe the radial pressure~(in all cases $b(r)$ is the shape function, while $R(r)$ is the Ricci scalar). In all the cases considered, we have been able to obtain exact wormhole solutions, definitely proving, with the help of particular examples, that the shape function fulfills all the corresponding constraints, as it should be. Moreover, the study of the first wormhole model, given by $P_{r} = -B b(r)^{u}/ \rho^{\alpha }$ for the radial pressure, has concluded that the NEC~(defined as $\rho + P_{r}$) in terms of $P_{r}$ can be violated at the throat of the wormhole. We have also seen, that it will be valid far from the throat. On the other hand, the NEC in terms of $P_{l}$, and the DEC in terms of the two pressures defined as $\rho - P_{r}$ and $\rho - P_{l}$, respectively, are always fulfilled. It should be mentioned that $\rho \geq 0$ proves that only the violation of the WEC in terms of the $P_{r}$-pressure will be observed at the throat of the wormhole, due to the violation of the NEC in terms of $P_{r}$. 

The study of $\omega_{r} = P_{r}/\rho$ has led to the conclusion that we should expect wormhole formation provided the radial equation of state parameter has a phantom behavior. From the study of the second wormhole model with $P_{r} = -B R(r)^{m}/ \rho^{\alpha }$, one can expect to find wormhole models for which only the NEC in terms of the $P_{l}$ and the DEC in terms of the $P_{r}$ pressures will be valid, at the throat of the wormhole. On the other hand, the NEC in terms of $P_{r}$ and the DEC in terms of $P_{l}$ are not valid at the throat. Far from the throat all mentioned energy conditions are fulfilled.  

A similar picture has been obtained from the study presented in the second part of the paper for the models described by varying barotropic fluids. Thus, in the case of the wormhole model obtained for $k=-5$ and $\eta = -2$, described by $P_{r} =\hat{ \omega} r^{k} R(r)^{\eta}  \rho$, we can claim that the NEC and the WEC in terms of $P_{r}$ are not fulfilled at the throat of the wormhole with $r_{0} = 1$. On the other hand, the DEC in terms of $P_{l}$ is also not valid at the throat. Similarly to the other models, the validity of all energy conditions in terms of both pressures has been observed for large values of $r$, i.e., far from the throat of the wormhole. 

We would also like to mention that the NEC and the WEC in terms of $P_{l}$, with the DEC in terms of $P_{r}$, are satisfied everywhere. Finally, we should note that the study of a particular wormhole model, given by $P_{r} = \omega b(r)^{v} \rho$, provides the same qualitative behavior for the energy conditions as in case of the model obtained for $P_{r} = -B b(r)^{u}/ \rho^{\alpha }$. Moreover, for the models with $P_{r} = -B b(r)^{u}/ \rho^{\alpha }$, and $P_{r} = \omega b(r)^{v} \rho$, we have studied the impact of the  parameters $u$ and $v$ on the energy conditions. We observed that, in both cases, there are some regions, corresponding to small $r$, where all the energy conditions are satisfied. On the other hand, the violation of the NEC in terms of $P_{r}$ will be observed for relatively small values of $r$, while the other energy conditions can still remain valid. 

A general conclusion, drawn from the study carried out in this paper, is that we can obtain exact wormhole models when the radial equation of the equation of state state parameter has phantom nature. On the other hand, since the equation of state of the wormhole is not well understood and constrained, we can try to extract additional information on the matter content using the behavior of the energy conditions at the throat of the wormhole and also far from it. The general feature of the models discussed here is the violation of the NEC in terms of $P_{r}$ at the throat of the wormhole. 

Moreover, the models could actually be distinguishable in future cosmological observations, due to the violation, for instance, of the DEC in terms of the $P_{l}$-pressure at the throat of the wormhole. The analysis presented in this paper is just a first, pioneering step towards the construction of more elaborated models of this type. As is obvious, in order to obtain exact solutions, we have  here restricted our attention to a very simple form of the $f(\textit{R}, \textit{T})$ theory of gravity. Therefore, one of the possible directions in future investigations will be the consideration of other forms in this class $f(\textit{R}, \textit{T})$. 

On the other hand, we expect to report, in future papers, on wormhole models constructed using more exotic forms of the varying Chaplygin gas. In the recent literature there is another non-linear equation of state, for what is termed a van der Waals fluid, which has been used to describe dark energy. Various studies point towards the conclusion that this dark energy representation can work well. Therefore, it seems reasonable to study wormhole formation with this matter content, as well. We expect to report some results on the studies of this direction soon. To finish, in all cases, the stability analysis of the corresponding models should be performed, too.

\section*{Acknowledgments}
EE has been partially supported by MINECO (Spain), Project FIS2016-76363-P, by the Catalan Government 2017-SGR-247, and by the CPAN Consolider Ingenio 2010 Project.
MK is supported in part by the Chinese Academy of Sciences President's International Fellowship Initiative Grant (No. 2018PM0054).
The authors are also thankful to an anonymous referee, whose valuable comments have helped to improve the final version of this paper.

\end{document}